\documentstyle[prl,aps]{revtex}

\def\ut#1{\lower1.2ex\hbox{$\mathchar"3218$}\mkern -14mu%
          \hbox to 2ex{\hss$#1$\hss}}

\begin{document}
\draft
\twocolumn[\hsize\textwidth\columnwidth\hsize\csname
@twocolumnfalse\endcsname

\title{Decoherence of Bose-Einstein condensates in traps at finite
temperature}
\author{Robert Graham}
\address{Fachbereich Physik, Universit\"at-Gesamthochschule Essen\\
45117 Essen\\ Germany}
\maketitle

\begin{abstract}
The phase diffusion of the order parameter of trapped Bose-Einstein
condensates at temperature $k_BT\gg\hbar\bar{\omega}$ is determined,
which gives
the fundamental limit of the line-width $\Delta\nu$ of an atom laser. In
addition a prediction of the number fluctuations in the condensate and
their correlation time $\tau_c$ is made and a general relation for
$\Delta\nu\,\tau_c$ is derived from the fluctuation-dissipation
relation.
\end{abstract}

\pacs{03.75.Fi, 05.30.Jp, 67.40.Db, 67.40.Fd}

\vskip2pc]
\narrowtext

Bose-Einstein condensation in a weakly interacting Bose-gas in three
dimensions in the thermodynamic limit of an infinitely extended system
is a
second order phase transition in which an order parameter, the
macroscopic
wave-function, appears spontaneously with a fixed but arbitrary phase,
turning the global $U(1)$-symmetry responsible for particle-number
conservation
into a broken or hidden symmetry \cite{-1}.
The rigidity of the phase of the order parameter against local
perturbations
and the absence of any phase diffusion gives rise to the Goldstone
modes,
which take the form of collisionless (zero)-sound or hydrodynamic sound,
respectively, depending on whether the sound frequency is in the
collisionless
mean-field regime or in the collision-dominated regime
\cite{-1},\cite{0}.

In finite systems, and thus also in all trapped Bose-gases, sharp
phase-transitions are impossible and hidden symmetries in a rigorous
sense
cannot
appear \cite{-1}. Nevertheless a macroscopic wave-function describing a
Bose-Einstein
condensate still exists, as is now firmly established by the experiments
\cite{1}.
However, for the general reason mentioned, the phase of the macroscopic
wave-function cannot be stable but must undergo a diffusion process,
which
restores the $U(1)$ symmetry over sufficiently long time intervals
\cite{-1}.
 This
diffusion process is therefore different from the Goldstone modes
mentioned
before, which are oscillations around a fixed value of the phase and do
not restore the symmetry \cite{-1},\cite{0}.

Recently a first attempt has been made to measure the stability of the
phase
of the macroscopic wave function
in a trapped BEC. In an experimental set-up of considerable ingenuity
\cite{2}
the relative phase of two BEC's was measured using a time-domain
separated-oscillatory-field condensate interferometer. Over the
time-interval
of 100 ms scanned in the experiment the relative phase was found to be
robust.
At first sight this experimental result may seem surprising since
decoherence of entangled states of many atoms should be extremely rapid.
Then,
however, one realizes that there is so far no clear theoretical
prediction of
the decoherence-time of Bose-Einstein condensates in traps against which
the
aforementioned experiment, or extensions of it which will surely follow,
could
be checked. In a number of papers \cite{3} the dispersion of the phase
of
a trapped Bose-Einstein condensate at zero-temperature was considered,
which
is due to thermodynamic fluctuations $\delta\mu$ of the chemical
potential
$\mu$ in a finite system with fixed particle number. An extension of
this
mechanism to finite temperature has also been proposed \cite{4}. This
effect
is not a `phase diffusion' but corresponds to an effect of inhomogeneous
broadening, and is even reversible in `revivals'. The experiments are
done at `high' temperature $k_BT\gg\hbar\bar{\omega}$ and even
$k_BT\gg\mu$, where
$\bar{\omega}$ is the geometrical mean of the three main trap
frequencies.
One would expect a proper
phase-diffusion process to occur in such a regime due to the interaction
of
the condensate with a thermal bath of collective modes and
quasiparticles
but so far an understanding of this process seems lacking (see however
 \cite{4a}).
This gap in our understanding of Bose-Einstein condensates in
traps at finite temperature is particularly painful, because the
fundamental
limit of the line-width of an atom laser depends on it: there is not yet
a `Schawlow-Townes' formula \cite{4b} for the linewidth of an atom
laser,
because its derivation requires a prior understanding of phase-diffusion
in
Bose-Einstein condensates.

In the present paper I intend to outline a theory of dissipation and
thermal
fluctuations in a trapped Bose-Einstein condensate which is used to
determine
the phase-diffusion constant, and from it, the line-width of a trapped
Bose-Einstein condensate as a function of temperature. The result
obtained
explains the experimentally observed robustness of the phase. I find it
convenient to present first a phenomenological framework for the theory,
in
the form of Langevin equations
in which dissipation appears via phenomenological parameters and the
fluctuation-dissipation relation is invoked to determine the
fluctuations.
Then the phenomenological parameters are fixed by drawing on known
microscopic
results for the damping of collective modes and calculating the new
transport
coefficient in the Langevin equation of the condensate. It is found to
result
primarily from scattering of thermally excited collective modes
(phonons) off the condensate.
This coefficient then determines the phase-diffusion
constant and the fundamental line-width of an atom laser via a
Schawlow-Townes-type formula.

The weakly interacting Bose-gas in a trap
in standard notation is described by the Hamiltonian
\begin{equation}
 H=\int\!d^3\!x
 \hat{\psi}^+\Big\{-\frac{\hbar^2}{2m}\nabla^2+
  V(\bbox{x})-\mu+\frac{U_0}{2}\hat{\psi}^+
  \hat{\psi}\Big\}\hat{\psi}\,.
 \label{eq:H}
\end{equation}
The total number of atoms $N$ is fixed. $\mu$ is the chemical potential.
The
presence of a Bose-Einstein condensate means that many ($N_0\gg1$)
particles occupy the normalized single-particle state $\psi_0(\bbox{x}$
of lowest
energy $\mu$ satisfying \cite{5}
$ -(\hbar^2/2m)\nabla^2\psi_0+
  \big(V(\bbox{x})+U_0N_0|\psi_0|^2\big)\psi_0=\mu\psi_0$.

The number density of the condensate is $n_0(\bbox{x})=
N_0|\psi_0(\bbox{x})|^2$. Both $\psi_0(\bbox{x})$ and $N_0$ are
functions of
$\mu$. In the following we turn this around and consider $\mu$ a
function of
$N_0$. The presence of the highly
occupied condensate mode makes  the decomposition of the Heisenberg
field-operator $\hat{\psi}(\bbox{x},t)=\big(\alpha_0(t)\psi_0(\bbox{x})+
\hat{\psi}'(\bbox{x},t)\big)e^{-i\mu t/\hbar}$ useful, where we shall
follow
the tradition starting with Bogoliubov \cite{7} and describe the
condensate
classically, remembering however, that
$N_0=|\alpha_0|^2$ is the particle number in the condensate.
$\hat{\psi}'(\bbox{x},t)$ is the
field operator for the particles outside the condensate. The Hamiltonian
then
splits according to $H=H_0+H_1+H_2+H_3+H_4$ where $H_n$ ($n=0,1,2,3,4$)
comprises the terms of $H$ which are of n-th order in $\hat{\psi}',
\hat{ \psi}^{'+}$, respectively. The equation of motion for the
condensate
amplitude $\alpha_0$ receives contributions from $H_0$ to $H_3$.
For a discussion of various microscopic approximation-schemes
see \cite{Griffin}.

Quasiparticle operators $\hat{a}_{\nu},\hat{a}_{\nu}^{+}$ are defined by
the standard Bogoliubov transformation $\hat{\psi}'(\bbox{x},t)=\sum_\nu
\big(u_\nu(\bbox{x})\hat{\alpha}_\nu(t)+
v^*_\nu(\bbox{x})\hat{\alpha}_\nu^+(t)\big)$,
where $u_\nu$, $v_\nu$ satisfy the usual Bogoliubov-De Gennes equations
\cite{Griffin}, and the $\hat{a}_{\nu}$ obey the
Heisenberg equations of motion
$\dot{\hat{\alpha}}_\nu=\frac{i}{\hbar}
[\hat{H}^{(\nu)},\hat{\alpha}_\nu]$
with
$\hat{H}^{(\nu)}=\hbar\omega_\nu(\hat{\alpha}_\nu^+\hat{\alpha}_\nu-\int|v_{\nu}(\bbox{x})|^2d^3x)$.  

Again $|u_{\nu}|^2$, $|v_{\nu}|^2$, and $\omega_{\nu}$ are functions of
$|\alpha_0|^2$.

In order to derive a phenomenological equation of motion for
the condensate alone we turn to the free energy $F(|\alpha_0|^2)$.
Expanded to
second order around its minimum it takes the form
\[
\beta
F(|\alpha_0|^2)=\frac{(|\alpha_0|^2-\langle|\alpha_0|^2\rangle)^2}{2\langle\delta  
N_0^2\rangle}
\]
 with $\langle|\alpha_0|^2\rangle=N-\langle\hat{N}'\rangle$,
$\langle\delta
N_0^2\rangle=\langle\hat{N'}^{2}\rangle-\langle\hat{N'}\rangle^2$,
where
$\hat{N'}=\int d^3x\hat{\psi'}^+(\bbox{x})\hat{\psi'}(\bbox{x})$. The
expectation values $\langle\hat{N'}\rangle$ and $\langle\delta
N_0^2\rangle$
(which turns out to be anomalously large $\sim N^{4/3}$) have recently
been
evaluated within the Bogoliubov theory \cite{11a} and can therefore here
be
considered as known. The equation of motion of $\alpha_0$ near thermal
equilibrium can now be written with the help of $F(|\alpha_0|^2)$ in
the
general form \cite{6a}
\begin{equation}
 i\hbar\dot{\alpha_0}=(1-i\Gamma_0)
  \frac{\partial F}{\partial\alpha_0^*}+F_0(t)
\label{eq:1}
\end{equation}
with Gaussian white noise $F_0(t)$ satisfying
\[
 \langle F_0(t)\rangle=0\,,\quad \langle F_0^*(t)F_0(0)\rangle=
  2\hbar k_BT\Gamma_0\delta(t)
\]
determined so as to ensure the correct equilibrium distribution \cite{8}
$\rho(\alpha_0,\alpha_0^*)=Z_0^{-1}\exp(-F(|\alpha_0|^2)/k_BT)$
for the condensate. Here I ignore the possibility
of the occurrence of squeezing in the thermal bath of uncondensed
particles.
Then
only a single new phenomenological coefficient $\Gamma_0$,
a dimensionless but
surely temperature dependent number, remains to be determined below.

As a short digression let us also extend the
equations of motion of the quasi-particles to include dissipation
and fluctuation within a phenomenological Markoffian framework \cite{9}.
It is convenient to do this by writing quantum Langevin equations
\cite{9a}
\begin{equation}
 i\hbar\dot{\hat{\alpha}}_\nu=(1-i\Gamma_\nu)
  \frac{\partial
\hat{H}^{(\nu)}}{\partial\hat{\alpha}_\nu^+}+\hat{F}_\nu(t)
\label{eq:*}
\end{equation}
with Gaussian Langevin-force operators satisfying $\langle
\hat{F}_\nu(t)\rangle = 0$,
$\langle [\hat{F}_\nu(t),\hat{F}_{\nu'}^+(t')]\rangle =
2\hbar^2\Gamma_\nu\omega_\nu\delta(t-t')\delta_{\nu\nu'}$ and, by the
fluctuation dissipation theorem,
\begin{equation}
\langle \hat {F}_\nu^+(t)\hat{F}_{\nu'}(t')\rangle =
2\hbar^2\Gamma_\nu\omega_\nu     \bar{n}_\nu\delta(t-t')\delta_{\nu\nu'}
\label{eq:twostars}
\end{equation}
with the Planck-distribution
$\bar{n}_\nu=(e^{\beta\hbar\omega_\nu}-1)^{-1}$.
The phenomenological
coefficients $\Gamma_\nu$ have the meaning of one half of the inverse
$Q$-factor of mode $\nu$, $\Gamma_\nu=(2Q_\nu)^{-1}$, and
remain to be determined below.

But let us now consider how number
fluctuations and the phase-diffusion of the condensate in equilibrium
follows
from (\ref{eq:1}). Its  deterministic part describes
the relaxation of the condensate to the minimum of the free energy
$F$ of the condensate at $\langle
N_0\rangle=\langle|\alpha_0|^2\rangle$. The
particle number fluctuations
$\delta N_0=|\alpha_0|^2-\langle|\alpha_0|^2\rangle$ in equilibrium,
after
linearizing (\ref{eq:1}) in $\delta N_0$, are found
to have the correlation function
\begin{equation}
 \langle\delta N_0(t)\delta N_0(t')\rangle = \langle\delta N_0^2\rangle
  e^{-|t-t'|/\tau_c}
\label{eq:3a}
\end{equation}
with the correlation time
$\tau_c=(\hbar \langle\delta N_0^2\rangle/2\Gamma_0\langle N_0\rangle
k_BT)$
which could in principle be measured
by taking the Fourier transformation in time of time-resolved in situ
phase-contrast images of the condensate. On a time-scale very much
larger
than the correlation time $\tau_c$ the phase $\varphi_0$ of the
condensate in equilibrium, i.e. the phase of
$\alpha_0=\sqrt{N_0}e^{i\varphi_0}$, satisfies
the Langevin equation of a Wiener process
with diffusion constant
\begin{equation}
 D_\varphi= k_BT(\Gamma_0 +\Gamma_0^{-1})/(\hbar \langle N_0\rangle),
\label{eq:PD}
\end{equation}
i.e $\langle\big(\varphi_0(t)-\varphi_0(0)\big)^2\rangle=D_\varphi t$.
The expectation value $\langle\alpha_0(t)\rangle$ then decays
exponentially
according to $\langle\alpha_0(t)\rangle=\sqrt{\langle
N_0\rangle}e^{-\Delta\nu t}$ with the
linewidth $\Delta\nu$ given by the Schawlow-Townes-type formula
\begin{equation}
 \Delta\nu=k_BT(\Gamma_0 +\Gamma_0^{-1})/(2\hbar \langle N_0\rangle)\ge
k_BT/(\hbar N_0)\,.
\label{eq:STH}
\end{equation}
The general relation between $\Delta\nu$ and $\tau_c$
\begin{equation}
 \Delta\nu=\frac{k_BT}{2\hbar\langle N_0\rangle}\big
(\frac{\hbar\langle\delta N_0^2\rangle}{2\langle N_0\rangle
k_BT\tau_c}+\frac{2\langle N_0\rangle k_BT\tau_c}{\hbar \langle\delta
N_0^2\rangle}\big )
\label{eq:5a}
\end{equation}
follows from our phenomenological theory, which is independent of the
yet
unknown coefficient $\Gamma_0$ and holds for general value of $k_BT$
(outside
the critical region). Eq.~(\ref{eq:5a}) is a general consequence of the
fluctuation dissipation relation for the condensate in the absence of
squeezing in the bath of uncondensed atoms.

Let us now determine the numbers $\Gamma_0$, $\Gamma_\nu$ from
microscopic
considerations, starting with $\Gamma_\nu$ for $\nu\ne 0$. We shall here
confine our attention to the damping of the low-lying collective modes
in the
collisionless regime, even though our phenomenological framework may
still be
used in the collision-dominated regime.
Furthermore, we confine ourselves to the experimentally relevant regime
$k_BT\gg\mu$. The damping of the low lying collective modes in this
regime
is due to Landau damping, described by a part of $H_3$, and was
calculated
for spatially homogeneous
condensates first by Sz\'epfalusy and Kondor \cite{10}. Their result
written
for our coeficient $\Gamma_\nu$ yields
 $\Gamma_\nu=A_\nu\ (k_BT/\mu)(n_0a^3)^{1/2}$.
The numerical coefficient $A_\nu$ turns out to be independent of $\nu$
in
the spatially homogeneous system, and its value there is \cite{11}
$A_\nu=3\pi^{3/2}/4$. However, a similar expression for $\Gamma_\nu$
was even shown to hold for the collective excitations in traps \cite{12}
where $n_0$ is the condensate density $n_0(\bbox{0})$
in the center of the trap and $A_\nu$
is a numerical coefficient which depends on the trap geometry and the
mode
function for mode $\nu$. For the dipole modes $A_\nu$ must vanish by the
Kohn-theorem \cite{K}.

The coefficient $\Gamma_0$ is unknown
so far and needs to be calculated from scratch. Here we shall determine
the temperature dependence of $\Gamma_0$ for large quasi-homogeneous
condensates for which the local density approximation is applicable,
collecting the trap-dependence of $\Gamma_0$ in a dimensionless
prefactor which we leave undetermined. Collisions of quasiparticles
with the condensate changing the
particle number $|\alpha_0|^2$ in the condensate by
$\Delta|\alpha_0|^2=\pm1$
are described by $H_3= \int d^3 xU_0(\alpha_0\psi_0(\hat{\psi}^{'+})^2
\hat{\psi}'+{\it h.c.})$. In such processes the
energy changes only by a tiny amount $\Delta F=
\partial F/\partial|\alpha_0|^2$ (of the order of
$k_BT/\sqrt{\langle\delta N_0^2\rangle}$). However,
processes described by $H_3$ involve three quasiparticles besides one
condensate particle, and they can therefore take up, from a slightly
perturbed condensate, an arbitrarily small amount $\Delta F$ of its
free
energy.

By the golden rule the rate $\gamma = d\langle
|\alpha_0|^2\rangle/dt$
is given by
\begin{eqnarray}
\gamma =
 -\frac{2\pi}{\hbar^2} &\displaystyle{\sum_{\nu,\mu,\kappa}}&
   \frac{1}{2}\langle |M^{(1)}_{\kappa,\nu\mu}|^2
    \delta(\omega_\kappa-\omega_\nu-\omega_\mu+\Delta
F/\hbar)\nonumber\\
   &&\label{A}-|M^{(2)}_{\nu\mu,\kappa|}|^2
    \delta(\omega_\kappa-\omega_\nu-\omega_\mu-\Delta F/\hbar)\rangle\\\
   &&[\bar{n}_\nu\bar{n}_\mu(\bar{n}_\kappa+1)-(\bar{n}_\nu+1)
    (\bar{n}_\mu+1)\bar{n}_\kappa]\nonumber
 \end{eqnarray}
The relevant matrix elements are
\begin{eqnarray*}
&M^{(1)}_{\kappa,\nu\mu}=2U_0\alpha_0\int d^3x\psi_0v_\nu(u^*_\kappa
u_\mu+\frac{1}{2}v^*_\kappa v_\mu)+(\nu \leftrightarrow\mu)\\
&M^{(2)}_{\nu\mu,\kappa}=2U_0\alpha_0\int d^3x\psi_0u^*_\nu (v^*_\mu
v_\kappa +\frac{1}{2}u^*_\mu u_\kappa)+(\nu \leftrightarrow\mu)
 \end{eqnarray*}
$M^{(1)}$ describes a Landau-scattering process
in which one atom is scattered out of the condensate by the absorption
of the two quasiparticles $\nu, \mu$ out of and the emission
of the new quasiparticle $\kappa$ into the thermal bath. Likewise
$M^{(2)}$ describes Beliaev scattering where an incoming thermal
quasiparticle
$\kappa$ is absorbed, an atom is kicked out of the condensate, and two
quasiparticles $\nu, \mu$ are emitted into the thermal bath. The factor
$1/2$ in (\ref{A}) accounts for the indistinguishability of
pairs $\nu, \mu$ and $\mu, \nu$.

In the phonon part of the excitation spectrum we have $u_\lambda\simeq
-v_\lambda\sim\omega^{-1/2}_\lambda$. Furthermore, in that low- energy
region
the statistical factor $[...]$ in (\ref{A}) is well approximated by
$(k_BT)^2(\omega_\kappa-\omega_\nu-\omega_\mu)/\hbar^2
\omega_\nu\omega_\mu\omega_\kappa$, where the frequency difference in
the nominator becomes $\pm\Delta F/\hbar$ in the product with the
$\delta$-functions, which express energy-conservation.
Anywhere else the small energy $\Delta F$ is negligible. The frequency
factors in the denominator, together with similar factors
in the denominator coming from the matrix elements,
make the phonon contribution to the sums in
(\ref{A}) the dominant one, at least in large condensates, and we shall
therefore concentrate on this contribution in the following.
This frequency range has a natural upper cut-off at $\mu/\hbar$, where
the collective phonons go over smoothly into particle-like
excitations. In finite condensates also a natural lower cut-off exists
at about the trap frequency $\bar{\omega}$, where the phonon wavelength
becomes comparable with the size of the condensate.

We shall evaluate the sums in (\ref{A}) in the usual Thomas-Fermi
and local density approximation \cite{rev} for $u_\lambda, v_\lambda,
\omega_\lambda$, integrating over the wave vectors of the phonons
in an interval implied by the cut-offs. We obtain in this way
\[
\gamma = -\frac{9A_0m^3(U_0k_BT)^2}{(2\pi)^3\hbar^8}\langle\Delta
F|\alpha_0|^2\rangle\int\int\frac{d\omega_\nu d\omega_\mu}
{\omega_\nu\omega_\mu(\omega_\nu+\omega_\mu)}
\]
Clearly the contribution near the lower cut-off at $\bar{\omega}$
dominates in the double frequency integral.
 After its evaluation, and using $U_0=4\pi\hbar^2 a/m$ and
$d_0=\sqrt{\hbar/m\bar{\omega}}$ to introduce the s-wave scattering
length
$a$ and the zero-point amplitude $d_0$, we obtain to leading order in
$\hbar\bar{\omega}/\mu$
$d\langle|\alpha_0|^2\rangle/dt=\gamma =-2(\Gamma_0/\hbar)
\langle(\partial F/\partial |\alpha_0|^2)|\alpha_0|^2\rangle$
with
\begin{equation}
 \Gamma_0=A_0\frac{18\ln2}{\pi}\big
(\frac{ak_BT}{d_0\hbar\bar{\omega}}\big )^2
\label{eq:R1}
\end{equation}
Here $A_0$ depends on the trap
geometry and is of order 1 in an isotropic trap. For the experimentally
realized condensates $\Gamma_0<<1$ is implied
by (\ref{eq:R1}).

In order to
give a practically useful estimate of the phase-diffusion rate let us
compare the line-width $\Delta\nu$ of the condensate with the directly
measurable inverse life-time $\tau_\kappa^{-1}$ of a collective mode
$\kappa$
with frequency of order $\bar{\omega}$. We find for $\Gamma_0 <<1$
\begin{equation}
 \Delta\nu\,\tau_\kappa=
\frac{\bar{\omega}/\omega_\kappa}{A_0A_\kappa 288\ln 2}
\big [\frac{(\mu/k_BT)^2}{\langle N_0\rangle (n_0(0)a^3)^{3/2}}\big ]\,.
\label{eq:R2}
\end{equation}

For $\mu/k_BT=10^{-1}, \langle N_0\rangle =10^6, n_0(0)a^3=3.10^{-6}$
the
factor $[...]$ on the
right hand side of (\ref{eq:R2}) is about 2. This may explain the
experimentally observed \cite{2} robustness of the phase.
Another way to put our result is to note that up to a numerical
prefactor
the relation $\hbar
\bar{\omega}\Gamma_0/\mu\sim\Gamma_\kappa^2=(2Q_\kappa)^{-2}$ holds.
Hence,
according to
(\ref{eq:STH}) for $\Gamma_0<<1$
\begin{equation}
\Delta\nu=(A^2_\kappa/A_0 36\ln 2)(k_BT\bar{\omega}Q_\kappa^2
/\mu\langle
N_0\rangle ) \label{eq:11}
\end{equation}
where $Q_\kappa$ is known experimentally.

In conclusion we have provided a theory of thermally generated
phase-diffusion
in Bose-Einstein condensates and determined a temperature dependent
Schawlow-Townes type formula for the linewidth $\Delta\nu$ of the
condensate.
This is also the minimum line-width of an atom laser based on the
Bose-Einstein condensate.
We have also
calculated the correlation time $\tau_c$ of the particle number
fluctuations in the condensate which should be measurable. Last not
least
we have derived a relation (\ref{eq:5a}) between this correlation time
$\tau_c$ and the line-width $\Delta\nu$ which follows from the
fluctuation-dissipation relation independently
of any microscopic detail, but subject to the assumption of negligible
squeezing in the thermal bath seen by the condensate. Removing this
restriction
will be the subject of a more detailed paper.

\section*{Acknowledgement:}

This work has been supported by the Deutsche Forschungsgemeinschaft
through
the Sonderforschungsbereich 237 ``Unordnung und gro{\ss}e
Fluktuationen''.
I would like to thank Dr. Walter Strunz for alerting me to an error in
an
earlier draft of this manuscript.


\begin{thebibliography}{99}
\bibitem{6a} It is not sufficient to use $H_0(|\alpha_0|^2)$ in place of
$F$
             as neither $\langle |\alpha_0|^2\rangle$ nor $\langle\delta
             N_0^2\rangle$ would then be given correctly for $T\ne0$.
\end{thebibliography}
\end{document}